\documentstyle[aaspp4,epsbox,11pt]{article}

\lefthead{Nishiura, Murayama, \& Taniguchi}
\righthead{BLRs and Accretion Disks}

\begin{document}
\title{A Geometrical Relationship between Broad-Line Clouds and 
an Accretion Disk around Active Galactic Nuclei}
%
%
\author{Shingo Nishiura, Takashi Murayama, and Yoshiaki Taniguchi}
\affil{Astronomical Institute, Tohoku University, Aoba, Sendai 980-77, Japan;
{\it E-mail:} nishiura@astroa.astr.tohoku.ac.jp}
%
%
\begin{abstract}

Recent hard X-ray spectroscopy of active galactic nuclei has strongly 
suggested that double-peaked, very broad Fe K emission arises from an 
accretion disk around the central engine. Model fitting of the observed 
Fe K emission line profile makes it possible to estimate a probable 
inclination angle of the accretion disk. In order to study the geometrical 
relationship between the accretion disk and broad emission-line regions 
(BLRs), we investigate the correlation between the inclination angle of 
the accretion disk and the velocity width of BLRs for 18 type-1 Seyfert 
galaxies. We found that there may be a negative correlation between them, 
i.e., Seyfert nuclei with a more face-on accretion disk tend to have larger 
BLR velocity widths, suggesting that the BLRs are not coplanar with respect 
to the accretion disk. The most probable interpretation may be that the BLRs 
arise from outer parts ({\it r} $\sim$ 0.01 pc) of a warped accretion disk 
illuminated by the central engine.
\end{abstract}
%
%
%
\section{Introduction}
Given the current paradigm of active galactic nuclei (AGNs), the observed 
huge luminosities of AGNs are powered by gravitational accretion onto 
a supermassive black hole (e.g., Rees 1984; Blandford 1990; Antonucci 1993; 
Peterson 1997). The central engines are considered to be surrounded by dusty 
tori whose typical inner radii are on the order of $\sim$ 1 pc (e.g., 
Antonucci, Miller 1985; Pier, Krolik 1992, 1993). Therefore, in order to 
understand AGNs, it is very important to investigate the spatial structures 
of the inner $\sim$ 1 pc regions in which a supermassive black hole, an 
accretion disk, warm absorbers, and broad emission-line regions (BLRs) reside. 

The innermost constituent is the accretion disk; its typical radius is $\sim$ 
(10 --- 100) $R_{\rm S} \sim (10^{-4}$ --- $10^{-3})M_{8}$ pc where 
$R_{\rm S}$ is the Schwarzschild radius of a black hole and $M_{8}$ is the 
black-hole mass in units of $10^{8} M_{\sun}$. The existence of accretion 
disks in AGNs has been demonstrated by the recent X-ray spectroscopy, 
collected using {\it ASCA} (Tanaka et al. 1994). The {\it ASCA} X-ray spectra 
of type-1 AGNs show the presence of very broad Fe K$\alpha$ emission, whose 
line profile can be fitted well by some accretion-disk models (Tanaka et al. 
1995; Fabian et al. 1995; Mushotzky et al. 1995; Iwasawa et al. 1996; Nandra 
et al. 1997; Reynolds 1997). An ionized accretion disk has also been detected 
by the recent radio continuum mapping in the archetypical, nearby Seyfert 
galaxy NGC 1068 (Gallimore et al. 1997). 

Another inner constituent is broad-line regions (BLRs), because their typical 
radii are of on the order of 0.01 pc (e.g., Peterson 1993). One of the most 
important questions related to the BLRs is how emission-line clouds are 
distributed around AGNs. Although there is still no definite consensus 
concerning the dynamical and spatial structure of BLRs, there are three 
alternative models: 1) the disk model (Shields 1979; see also Osterbrock 
1989), 2) the high-velocity streamer model (Zheng et al. 1991), and 3) a pair 
of conical regions in which photoionized clouds are orbiting randomly with 
Keplerian motion (Wanders et al. 1995; Wanders, Peterson 1996, 1997; Goad, 
Wanders 1996). In particular, a recent detailed analysis of the reverberation 
mapping has strongly suggested that the BLRs of the type-1 Seyfert galaxy 
NGC 5548 has the third type of geometry (Wanders et al. 1995). It is, 
however, still not known which model is the most popular one. 

Since recent observations have shown that the BLRs are dominated by 
rotational motion, rather than the radial motion (e.g., Peterson 1993; 
Wanders et al. 1995), it is interesting to examine whether or not the 
rotational axis of the BLR is nearly the same as that of the accretion disk. 
If the disk model for BLRs would be correct, the BLRs may be coplanar with 
respect to the accretion disk. In fact, double-peaked BLRs have sometimes 
been considered to arise from accretion disks, themselves [e.g., P\'erez 
et al. (1988); Livio, Xu (1997) and references therein; see also, however, 
Gaskell (1996)]. Motivated by this, we investigate the relationship between 
the inclination angle of the accretion disk and the width of BLR 
statistically using published data. 
%
%
\section{Data}
Nandra et al. (1997) presented a systematic analysis of the {\it ASCA} X-ray 
spectra of 18 type-1 Seyfert galaxies. They fitted the Fe K$\alpha$ emission 
profiles and derived the most probable inclination angles for the following 
three models: Model A, the Schwarzschild model with reflection; Model B, the 
Schwarzschild model with the emissivity law for $q = 2.5$ (see below); and 
Model C, the Kerr model with reflection, where the Schwarzschild and Kerr 
models refer to those proposed by Fabian et al. (1989) and Laor (1991), 
respectively. In their model-fitting procedures, there are five parameters: 
1) the inner radius of the disk ($R_{\rm i}$), the outer radius 
($R_{\rm o}$), the inclination of the axis of rotation with respect to the 
line of sight ($i_{\rm AD}$), the rest energy of Fe K$\alpha$ line 
($E_{\rm K\alpha}$), and the line normalization ($I_{\rm K\alpha}$). They 
adopted $R_{\rm i} = 3 R_{\rm S}$, $R_{\rm o} = 500 R_{\rm S}$, and 
$E_{\rm K\alpha} = 6.4$ keV for Models A and B, while $R_{\rm i} = 
0.615 R_{\rm S}$ and $R_{\rm o} = 200 R_{\rm S}$ for Model C. In addition to 
the above parameters, the line emissivity is also another parameter, which 
is usually parameterized by a power law as a function of the radius; 
i.e., $R^{-q}$. Model B is the case for {\it q} = 2.5, which is an 
averaged value for the case of free line emissivity [see table 4 and 
figure 6 in Nandra et al. (1997)]. The inclination angles for the three 
models are summarized in table 1. Note that some Seyfert nuclei were observed 
by {\it ASCA} more than once (e.g., NGC 4151). In these cases, we tabulated 
the estimated inclination angles for all of the observations. 

We have compiled the line widths of both H$\alpha$ and H$\beta$ emission, 
FWZI (full width at zero intensity), for the above 18 Seyfert galaxies based 
on the literature. Since it is known that the BLR emission generally shows 
time variations (e.g., Peterson 1993), it would be desirable to obtain the 
data of both BLR and Fe K$\alpha$ emission simultaneously. However, all of the 
BLR data were taken far before the X-ray observations. In order to minimize 
any possible effect of time variations, we compiled more than-one 
measurements for each galaxy as much as possible, and then used the averaged 
value in a later analysis. For some Seyferts, FWHMs (full width at half 
maxima) of the emission lines are only available in the literature. In these 
cases, we estimated FWZIs using a relation FWZI = $2\sqrt3$ FWHM (Wandel, Yahil 
1985). The compiled data are summarized in table 2.
%
%
\section{Results}
In figure. 1, we compare sin $i_{\rm AD}$ with FWZI(H$\alpha$)/
$L^{1/4}_{\rm X}$ ({\it left column}) and with FWZI(H$\beta$)/
$L^{1/4}_{\rm X}$ ({\it right column}) for the three accretion-disk 
models A, B, and C, where $L_{\rm X}$ is the X-ray (2 --- 10 keV) luminosity. 
In these comparisons, we used both FWZI(H$\alpha$)/$L^{1/4}_{\rm X}$ and 
FWZI(H$\beta$)/$L^{1/4}_{\rm X}$, instead of their raw FWZI values. 
The reason for this is that the velocity width of BLRs depends on the 
central mass, even if the inner radius of the BLRs would be the same among 
the Seyferts, while the inclination angle of an accretion disk has no 
dependence on the black-hole mass. If the Eddington ratios are not very 
different among the Seyferts (i.e., the mass-to-luminosity ratios are nearly 
the same among the Seyferts), we expect the relationship FWZI $\propto 
L^{1/4}$, which is analogous to the so-called Faber-Jackson relation 
(Faber, Jackson 1976). We adopted the X-ray luminosities given by Nandra 
et al. (1997). 

Figure 1 shows that {\it all of the correlations are negative}. As shown in 
table 3, the correlation coefficients suggest that the negative correlations 
are not significant statistically. Therefore, our modest conclusion may be 
that there is a slight correlation between $\sin i_{\rm AD}$ with 
FWZI(H$\alpha$)/$L^{1/4}_{\rm X}$, suggesting a random orientation between 
the BLRs and the accretion disks. However, it is quite unlikely that the 
large errors in the estimate of $i_{\rm AD}$ cause any accidental negative 
correlations for all of the cases shown in figure 1. Therefore, we may 
conclude that there is a tendency of a negative correlation between 
$\sin i_{\rm AD}$ With FWZI(H$\alpha$)/$L^{1/4}_{\rm X}$. More interestingly, 
we mention that there is no hint of a positive correlation between them. 
%
%
\section{Discussion}
The most important result in this study is that there is no obvious 
{\it positive} correlation between $\sin i_{\rm AD}$ with FWZI(H$\alpha$)/
$L^{1/4}_{\rm X}$. This suggests that {\it the BLRs are not coplanar with 
respect to accretian disks}. 

Some Seyfert nuclei in our sample show double-peaked BLRs (DBLRs). It has 
sometimes been considered that such DBLRs may arise from an accretion disk, 
itself (e.g., P\'erez et al. 1988). The DBLR emission profiles of the four 
Seyfert nuclei in our sample (NGC 3227, NGC 3783, NGC 5548, and 3C 120) were 
studied by Rokaki et al. (1992) using a standard geometrically-thin accretion-
disk model; also, the inclination angles of the DBLRs ($i_{\rm BLR}$) were 
derived. We compare these inclination angles with those of the accretion 
disks in figure 2. This comparison also suggests a negative correlation 
between $i_{\rm AD}$ and $i_{\rm BLR}$, being consistent with our result. 
This strengthens our suggestion that the BLRs are not coplanar with respect 
to the accretion disks in the Seyfert nuclei studied here.

Let us consider what kind of geometrical configuration can explain the 
non-coplanar property. The negative correlation means that the normalized 
velocity width increases with decreasing inclination angle; i.e., {\it Seyfert 
nuclei with a more face-on accretion disk tend to have larger BLR velocity 
widths}. There may be three alternative ideas to explain this property. One 
is the bipolar streamer model (e.g., Zheng et al. 1990). If we observe the 
accretion disk from a face-on view, the velocity width would be widest because 
the bipolar wind flows along our line of sight. However, this model has an 
intrinsic difficulty, as claimed by Livio and Xu (1996), because the emitting 
region on the receding flow (jet) is obscured from view by the accretion disk; 
the standard, optically thick accretion disk is opaque up to $\sim$ 1 pc, and, 
thus, the BLR component behind the disk cannot be seen, because the typical 
radial distance of BLRs from the central engine is on the order of 0.01 pc 
(e.g., Peterson 1993). The second idea is that BLRs are located in nearly the 
same plane as that of an accretion disk, but are orbiting with poloar orbits. 
If a two-sided jet is ejected with a highly inclined angle with respect to 
the {\it global} accretion disk, we can explain the negative correlation. 
Such a jet model is briefly described by Norman and Miley (1984). This idea 
is consistent with the recent reverberation mapping result for the BLRs of 
NGC 5548 because the most likely geometry of the BLRs of this galaxy is a 
pair of conical regions in which photoionized clouds are orbiting randomly 
with Keplerian motion (Wanders et al. 1995; Wanders, Peterson 1996, 1997; see 
also Goad, Wanders 1996). This model may also have the same obscuration 
problem as that for the above streamer model. However, if the BLR clouds are 
moving at randomly oriented orbits (Wanders et al. 1995), there may be no 
obscuration problem. The third idea is that BLRs arise from outer parts of 
a warped accretion disk. The disk model for BLR is the standard idea (Shields 
1977; see also for a review Osterbrock 1989). It has been recently shown 
that accreting gas clouds probed by water-vapor maser emission at 22 GHz 
show evidence of significant warping (Miyoshi et al. 1995; Begelman, 
Bland-Hawthorn 1997). The warping of accretion disks can be driven by the 
effect of the radiation-pressure force (Pringle 1996, 1997). For typical AGN, 
the warping may occur at {\it r} $>$ 0.01 pc (Pringle 1997), which is at a 
similar distance as BLRs. Therefore, the warped-disk model can explain the 
observed negative correlation reasonably well. This model is schematically 
shown in figure 3. Since the degree of warping and the viewing angle are 
different from AGN to AGN, the negative correlation between $i_{\rm AD}$ 
and $i_{\rm BLR}$ may be blurred as obtained in our analysis, although 
the poor correlation may be also due to the large errors in the estimate of 
$i_{\rm AD}$.

\acknowledgements

We would like to thank Kazushi Iwasawa, Toru Yamada, and Youichi Ohyama for 
their useful discussion and comments. We would also like to thank the 
anonymous referee for his/her very useful comments and suggestions. TM was 
supported by the Grant-in-Aid for JSPS Fellows by the Ministry of Education, 
Science, Sports and Culture. This work was supported in part by the Ministry 
of Education, Science, Sports and Culture (No. 07044054).
%
%

%
%
%
%
%
%
%
\clearpage
\begin{table}[h]
\begin{center}
\caption{The X-ray data of type 1 seyfert galaxies.}
\vspace{10pt}
\begin{tabular}{lcccc}
\hline\hline
       & $\log L_{\rm X}$ (2---10 keV) & &{\it i}$_{AD}$ (degree) & \\
Object & (erg s$^{-1}$)                & Model A & Medel B & Model C \\
\hline      
Mrk 335        & 43.42 & 22$^{+68}_{-22}$ & 24$^{+13}_{-24}$ 
                        & 22$^{+68}_{-22}$ \\
Fairall 9      & 44.26 & 46$^{+44}_{-27}$ & 32$^{+12}_{ -8}$ 
                        & 89$^{ +1}_{-49}$ \\
3C 120         & 44.34 & 60$^{+30}_{-13}$ & 59$^{+10}_{-13}$ 
                        & 88$^{ +2}_{ -1}$ \\
NGC 3227       & 42.01 & 20$^{ +8}_{-14}$ & 23$^{ +8}_{ -6}$ 
                        & 21$^{ +7}_{-21}$ \\
NGC 3516       & 43.43 & 27$^{ +6}_{ -7}$ & 26$^{ +4}_{ -4}$ 
                        & 26$^{ +3}_{ -4}$ \\
NGC 3783(1)    & 43.25 & 35$^{ +2}_{-18}$ & 21$^{ +5}_{ -5}$ 
                        & 26$^{ +5}_{ -7}$ \\
NGC 3783(2)    & 43.25 & 32$^{ +3}_{-15}$ &  9$^{+11}_{ -9}$ 
                        & 40$^{+12}_{-40}$ \\
NGC 4051       & 41.56 & 34$^{ +3}_{-14}$ & 27$^{ +7}_{-11}$ 
                        & 25$^{+12}_{ -4}$ \\
NGC 4151(2)    & 42.97 & 17$^{+12}_{-17}$ & 20$^{ +5}_{ -5}$ 
                        &  9$^{+18}_{ -9}$ \\
NGC 4151(4)    & 42.97 & 33$^{ +2}_{-18}$ & 21$^{ +5}_{ -6}$ 
                        & 24$^{ +5}_{ -7}$ \\
NGC 4151(5)    & 42.97 & 17$^{ +5}_{ -5}$ & 15$^{ +4}_{ -5}$ 
                        & 21$^{ +5}_{-11}$ \\
Mrk 766        & 43.08 & 34$^{ +3}_{ -3}$ & 35$^{ +5}_{ -5}$ 
                        & 36$^{ +8}_{ -7}$ \\
NGC 4593       & 43.06 &  0$^{+79}_{ -0}$ & 45$^{+12}_{-11}$ 
                        &  0$^{+56}_{ -0}$ \\
MCG $-$6---30---15(1)& 43.07 & 33$^{ +3}_{ -5}$ & 34$^{ +5}_{ -5}$ 
                        & 34$^{ +5}_{ -6}$ \\
MCG $-$6---30---15(2)& 43.07 & 34$^{ +3}_{ -6}$ & 33$^{ +8}_{-25}$ 
                        & 34$^{+16}_{ -9}$ \\
IC 4329A       & 43.94 & 17$^{+14}_{-17}$ & 26$^{ +7}_{ -8}$ 
                        & 10$^{+13}_{-10}$ \\
NGC 5548       & 43.76 &  0$^{+76}_{ -0}$ & 39$^{+10}_{-11}$ 
                        & 10$^{+80}_{-10}$ \\
Mrk 841(1)     & 43.82 & 27$^{ +7}_{ -9}$ & 27$^{ +6}_{ -6}$ 
                        & 26$^{ +8}_{ -5}$ \\
Mrk 841(2)     & 43.82 & 38$^{ +2}_{-16}$ & 30$^{ +9}_{-15}$ 
                        & 90$^{ +0}_{-90}$ \\
Mrk 509        & 44.38 & 27$^{+63}_{-27}$ & 40$^{+48}_{-24}$ 
                        & 89$^{ +1}_{-89}$ \\
NGC 7469(2)    & 43.60 &  0$^{+89}_{ -0}$ & 45$^{+11}_{-29}$ 
                        & 20$^{+70}_{-20}$ \\
MCG $-$2---5---22   & 44.14 & 46$^{+44}_{-46}$ & 41$^{+10}_{-15}$ 
                        & 26$^{+64}_{-26}$ \\
\hline
\end{tabular}
\end{center}
\end{table}
%
%
\clearpage
\begin{table}[h]
\begin{center}
\caption{The optical data of type 1 Seyfert galaxies.}
\vspace{10pt}
\hspace*{-1.5cm}
\begin{tabular}{lcccccccc}
\hline\hline
       &      & {\it i$_{\rm BLR}$} & FWZI(H$\alpha$) &   &   
                                    & FWZI(H$\beta$)  &   &  \\
Object & Type & (degree)            & (km s$^{-1}$)   & Ref.$^{\#}$ 
       & adopt$^{\dag}$     & (km s$^{-1}$)   & Ref.$^{\#}$ 
       & adopt$^{\dag}$ \\
\hline
Mrk 335     & 1.0 &    &   9500 & 1  & 13620 &  10500 & 1  & 11967 \\
            &     &    &   9560 & 2  &       &  12800 & 3  &       \\
            &     &    &  21800 & 4  &       &  12600 & 5  &       \\
Fairall 9   & 1.0 &    &  16000 & 6  & 14450 &  10650 & 7  & 11025 \\
            &     &    &  12900 & 8  &       &  11400 & 8  &       \\
3C 120      & 1.0 & 20$^{\ast}$ &   9300 & 8  & 18900 &  15000 & 9  & 10667 \\
            &     &    &  28500 & 4  &       &   9900 & 8  &       \\
            &     &    &        &    &       &   7100 & 10 &       \\
NGC 3227    & 1.5 & 40$^{\ast}$ &  11400 & 1  & 16500 &  11800 & 1  & 11650 \\
            &     &    &  21600 & 4  &       &  11500 & 5  &       \\
NGC 3516    & 1.5 &    &  15500 & 1  & 15500 &  14800 & 1  & 13900 \\
            &     &    &        &    &       &  13000 & 5  &       \\
NGC 3783    & 1.2 & 25$^{\ast}$ &   9900 & 6  &  9900 &   9900 & 9  &  9900 \\
NGC 4051    & 1.0 &    &   6070 & 2  &  6070 &  10800 & 5  & 10800 \\
NGC 4151    & 1.5 &    &   9000 & 6  &  9000 &   9000 & 9  & 10000 \\
            &     &    &        &    &       &  11000 & 5  &       \\
Mrk 766     & 1.5 &    &   3118$^{\ddag}$ & 11 &  3118 &   7200 & 5  &  7200 \\
            &     &    &        &    &       &   7000 & 11 &       \\
NGC 4593    & 1.0 &    &  14000 & 6  & 14000 &  14549$^{\ddag}$ & 12 & 17205 \\
            &     &    &        &    &       &  20092$^{\ddag}$ & 13 &       \\
            &     &    &        &    &       &  16974$^{\ddag}$ & 14 &       \\
MCG $-$6---30---15& 1.0 &    &   5196$^{\ddag}$ & 15 &  5196 &   6500 & 6  &  6500 \\
IC 4329A    & 1.0 &    &  15600 & 6  & 22800 &  22000 & 16 & 22000 \\
            &     &    &  30000 & 16 &       &        &        &       \\
NGC 5548    & 1.5 & 50$^{\ast}$ &  14900 & 1  & 14900 &  15400 & 1  & 13300 \\
            &     &    &        &    &       &  11200 & 5  &       \\
Mrk 841     & 1.5 &    &  23500 & 17 &       &  11700 & 5  & 11700 \\
Mrk 509     & 1.2 &    &  12900 & 1  & 11700 &  15100 & 1  & 13100 \\
            &     &    &  10500 & 9  &       &  11100 & 8  &       \\
NGC 7469    & 1.0 &    &   7800 & 1  &  7800 &   8500 & 1  &  9900 \\
            &     &    &        &    &       &  11300 & 5  &       \\
MCG $-$2---58---22& 1.2 &    &  23700 & 8  & 24750 &  22700 & 7  & 19150 \\
              &     &    &  25800 & 2  &       &  15600 & 8  &       \\
\hline
\end{tabular}\vspace{0.5cm}\\
\end{center}
$^{\ast}$ Rokaki et al. (1992)
\par\noindent
$^{\dag}$ A simple average value is adopted.
\par\noindent
$^{\ddag}$ Estimated by FWZI = 2$\sqrt{3}$ FWHM (Wandel, Yahil 1985).
\par\noindent
$^{\#}$ References, 
(1) Osterbrock 1977;
(2) Osterbrock, Shuder 1982;
(3) Padovani, Rafanelli 1988;
(4) van Groningen, van Weeren 1989;
(5) Padovani et al. 1990;
(6) Steiner 1981;
(7) Ward et al.\ 1978;
(8) Rafanelli 1985;
(9) Elvis et al.\ 1978;
(10) Hooimeyer et al.\ 1992;
(11) Osterbrock, Pogge 1985;
(12) MacAlpine et al. 1979;
(13) Peterson et al.\ 1982;
(14) Crenshaw 1986;
(15) Pineda et al.\ 1980;
(16) Marziani et al. 1992;
(17) Eracleous, Halpern 1993.
\end{table}
%
%
\clearpage
\begin{table}[h]
\begin{center}
\caption{The correlation coefficients.}
\vspace{10pt}
\begin{tabular}{ccc}
\hline\hline
Model & FWZI(H$\alpha$)/$L_{\rm X}^{0.25}$ 
      & FWZI(H$\beta$)/$L_{\rm X}^{0.25}$ \\
\hline
Model A & $-0.125$ & $-0.360$ \\
Model B & $-0.239$ & $-0.331$ \\
Model C & $-0.306$ & $-0.638$ \\
\hline
\end{tabular}
\end{center}
\end{table}
%
%
\clearpage
\begin{figure}[hc]
\begin{center}
\epsfile{file=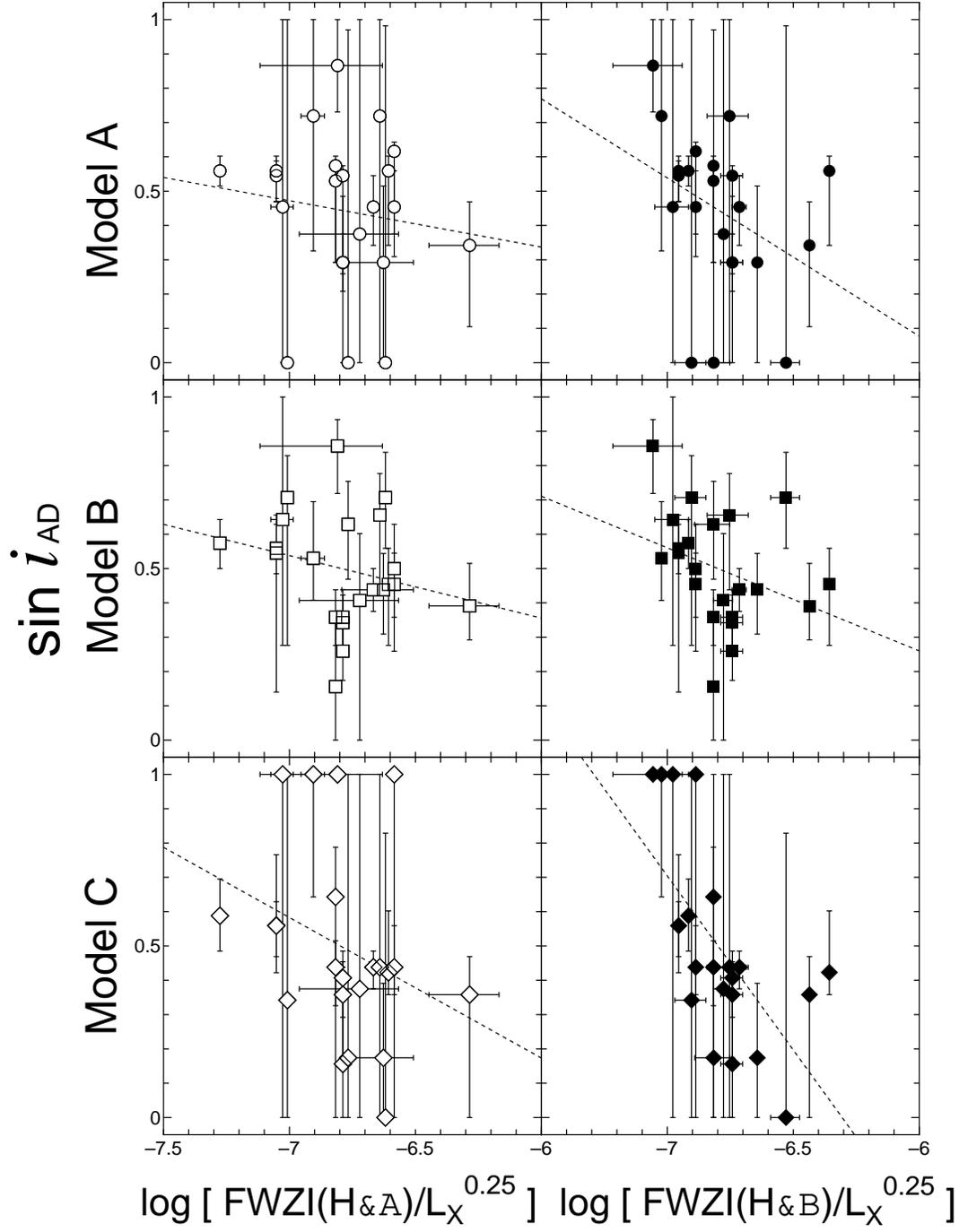}
\end{center}
\caption{Comparison of $\sin i_{\rm AD}$ with FWZI(H$\alpha$)/$L_{\rm X}$
({\it left panels}) and with FWZI(H$\beta$)/$L_{\rm X}$({\it right panels}).}
\end{figure}
%
%
\clearpage
\begin{figure}[hc]
\begin{center}
\epsfile{file=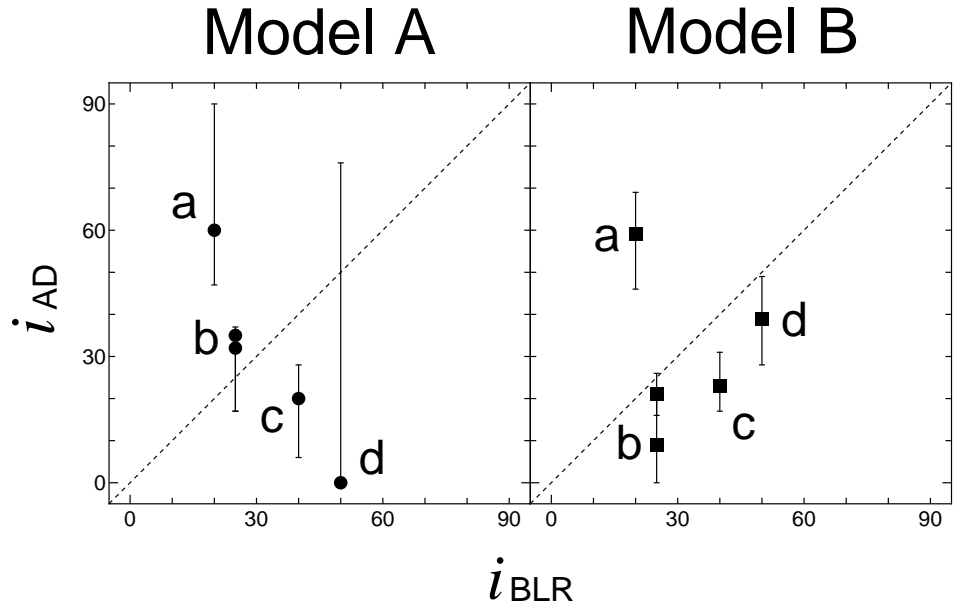}
\end{center}
\caption{Comparison between $\sin i_{\rm AD}$ and $\sin i_{\rm BLR}$ studied by 
Rokaki et al. (1992) for the following four Seyfert nuclei: a, 3C 120; 
b, NGC 3783; c, NGC 3227; d, NGC 5548.}
\end{figure}
\vspace{7cm}
%
%
%
\figcaption{Schematic illustration of the possible geometrical configuration between 
the Fe K emitting region ($r < 0.001$ pc) and the BLR ($r > 0.01$ pc) 
in a warped accretion disk.}

%

\clearpage
\begin{references}
\reference{1}{Antonucci R.R.J. 1993, ARA\&A, 31, 473}
\reference{1}{Antonucci R.R.J., Miller J.S. 1985, ApJ, 297,621}
\reference{1}{Begelman M.A., Bland-Hawthorn J. 1997, Nature, 385, 22}
\reference{1}{Blandford R.D. 1990, Active Galactic Nuclei, ed.  R.D. Blandford,
              H. Netzer, A. Woltier (Springer-Verlag), p. 161}
\reference{1}{Crenshaw D.M. 1986, ApJS, 62, 821}
\reference{1}{Elvis M., Maccacaro T., Wilson A.S., Ward M.J., Penston M.V.,
              Fosbury R.A.E., Perola G.G. 1978, MNRAS, 183, 129}
\reference{1}{Eracleous M., Halpern J.P. 1993, ApJ, 409, 584}
\reference{1}{Faber S.M., Jackson R.E. 1976, ApJ, 204, 668}
\reference{1}{Fabian A.C., Nandra K., Reynolds C.S., Brandt W.N., Otani C.,
              Tanaka Y., Inoue H., Iwasawa K. 1995, MNRAS, 277, L11}
\reference{1}{Fabian A.C., Rees M.J., Stella L., White N.E. 1989,
              MNRAS, 238, 729}
\reference{1}{Gallimore J.F., Baum S.A., O'Dea P. 1997, Nature, 388, 352}
\reference{1}{Gaskell C.M. 1996, in Jets from Stars and Galaxies, ed. W. Kundt
              (Springer, Berlin), p. 165}
\reference{1}{Goad M., Wanders I. 1996, ApJ, 469, 113}
\reference{1}{Hooimeyer J.R.A., Miley G.K., de Waard G.J., Schilizzi R.T.
              1992, A\&A, 261, 9}
\reference{1}{Iwasawa K., Fabian A.C., Reynolds C.S., Nandra K., 
              Otani C., Inoue H., Hayashida K., Brandt W.N., 
              Dotani T., Kunieda H., Matsuoka M., and Tanaka Y. 1996, 
              MNRAS, 282, 1038}
\reference{1}{Laor A. 1991, ApJ, 376, 90}
\reference{1}{Livio M., Xu C. 1997, ApJ, 478, L63}
\reference{1}{MacAlpine G.M., Williams G.A., Lewis D.W. 1979, PASP, 91, 746}
\reference{1}{Mannucci F., Salvati M., Stanga R.M. 1992, ApJ, 394, 98}
\reference{1}{Marziani P., Calvani M., Sulentic J.W. 1992, ApJ, 393, 658}
\reference{1}{Miyoshi M., Moran J., Herrnstein J., Greenhill L., Nakai N.,
              Diamond P., Inoue M. 1995, Nature, 373, 127}
\reference{1}{Mushotzky R.F., Fabian A.C., Iwasawa K., Matsuoka M., Nandra K.,
              Tanaka Y. 1995, MNRAS, 272, L9}
\reference{1}{Nandra K., George I.M., Mushotzky R.F., Turner T.J.,
              Yaqoob T. 1997, ApJ, 477, 602}
\reference{1}{Norman C.A., Miley G. 1984, A\&A, 141, 85}
\reference{1}{Osterbrock D.E. 1977, ApJ, 215, 733}
\reference{1}{Osterbrock D.E. 1989, 
              Astrophysics of Gaseous Nebulae and Active Galactic Nuclei
              ()University Science Books, California), p. 368}
\reference{1}{Osterbrock D.E. Pogge R.W. 1985, ApJ, 297, 166}
\reference{1}{Osterbrock D.E. Shuder J.M. 1982, ApJS, 49, 149}
\reference{1}{Padovani P., Burg R., Edelson R.A. 1990, ApJ, 353, 438}
\reference{1}{Padovani P., Rafanelli P. 1988, A\&A, 205, 53}
\reference{1}{P\'erez E., Mediavilla E., Penston M.V., Tadhunter C.,
              Moles M. 1988, MNRAS, 230, 353P}
\reference{1}{Peterson B.M. 1993, PASP, 105, 247}
\reference{1}{Peterson B.M. 1997, An Introduction to Active Galactic Nuclei
              (Cambridge University Press, Cambridge), p. 32}
\reference{1}{Peterson B.M., Foltz C.B., Byard P.L., Wagner R.M. 1982,
              ApJS, 49, 469}
\reference{1}{Pier E.A., Krolik J.H. 1992, ApJ, 401, 99}
\reference{1}{Pier E.A., Krolik J.H. 1993, ApJ, 418, 673}
\reference{1}{Pineda F.J., Delvaille J.P., Grindlay J.E., Schnopper H.W.
              1980, ApJ, 237, 414}
\reference{1}{Pringle J.E. 1996, MNRAS, 281, 357}
\reference{1}{Pringle J.E. 1997, MNRAS, 292, 136}
\reference{1}{Rafanelli P. 1985, A\&A, 146, 17}
\reference{1}{Rees M.J. 1984, ARA\&A, 22, 471}
\reference{1}{Reynolds C.S. 1997, MNRAS, 286, 513}
\reference{1}{Rokaki E., Boisson C., Collin-Souffrin S. 1992, A\&A, 253, 57}
\reference{1}{Shields, G. A. 1977, Ap. Lett., 18, 119}
\reference{1}{Steiner J.E. 1981, ApJ, 250, 469}
\reference{1}{Tanaka Y., Holt S.S., Inoue H. 1994, PASJ, 46, L37}
\reference{1}{Tanaka Y., Nandra K., Fabian A.C., Inoue H., Otani C., Dotani T., 
              Hayashida K., Iwasawa K., Kii T., Kunieda H., Makino F., and 
              Matsuoka M. et al. 1995, Nature, 375, 659}
\reference{1}{van Groningen E., van Weeren N. 1989, A\&A, 211, 318}
\reference{1}{Wanders I., Peterson B.M. 1996, ApJ, 466, 174}
\reference{1}{Wanders I., Peterson B.M. 1997, ApJ, 477, 990}
\reference{1}{Wanders I., Goad M., Korista K., Peterson B.M., Horne K., 
              Ferland G., Koratkar A., Pogge R.W., and Shields J. 1995, 
              ApJ, 453, L87}
\reference{1}{Wandel A., Yahil A. 1985, ApJ, 295, L1}
\reference{1}{Ward M.J., Wilson A.S., Penston M.V., Elvis E., Maccacaro T.,
              Tritton K.P. 1978, ApJ, 223, 788}
\reference{1}{Zheng W., Binette L., Sulentic J.W. 1990, ApJ, 365, 115}
\end{references}
\end{document}